\documentclass[aps,notitlepage,twocolumn]{revtex4-1}

\usepackage{graphicx}
\usepackage{amsmath}
\usepackage{amssymb}
\usepackage{graphics}

\newcommand{\cG}{{\cal G}}
\newcommand{\eg}{{e.g., }}
\newcommand{\ie}{{i.e., }}
\newcommand{\kT}{k_{\rm B}T}
\newcommand{\pd}{\partial}
\newcommand{\tencG}{{\boldsymbol{\cal G}}}

\newcommand{\vecf}{{\bf f}}

\newcommand{\vecP}{{\bf P}}
\newcommand{\vecq}{{\bf q}}
\newcommand{\vecr}{{\bf r}}
\newcommand{\vecR}{{\bf R}}
\newcommand{\vecu}{{\bf u}}
\newcommand{\vecv}{{\bf v}}
\newcommand{\xhat}{\hat{\bf x}}
\newcommand{\yhat}{\hat{\bf y}}
\newcommand{\lc}{\ell_c}
\newcommand{\Dbar}{\bar{D}}

\begin{document}

%========================================================================
\title{Correlations in suspensions confined between viscoelastic
  surfaces: \\ Noncontact microrheology}
%========================================================================

\author{Chen Bar-Haim}
\email{chenbar@tau.ac.il}
\affiliation{Raymond \& Beverly Sackler School of Chemistry, Tel Aviv
University, Tel Aviv 6997801, Israel}

\author{Haim Diamant} 
\email{hdiamant@tau.ac.il} 
\affiliation{Raymond \& Beverly Sackler School of Chemistry, Tel Aviv
University, Tel Aviv 6997801, Israel}

\date{September 13, 2017}

\begin{abstract}
We study theoretically the velocity cross-correlations of a viscous
fluid confined in a slit between two viscoelastic media. We analyze
the effect of these correlations on the motions of particles suspended
in the fluid. The compliance of the confining boundaries gives rise to
a long-ranged pair correlation, decaying only as $1/r$ with the
interparticle distance $r$. We show how this long-ranged effect may be
used to extract the viscoelastic properties of the confining media
without embedding tracer particles in them. We discuss the remarkable
robustness of such a potential technique with respect to details of
the confinement, and its expected statistical advantages over standard
two-point microrheology.
\end{abstract}

\maketitle
%------------------------------------------------

%% Introduction: Motivation. Relevant works on confined suspensions,
%% especially with long-ranged correlations. When screened and when
%% not. Momentum conservation. Rigid boundaries; relaxation of various
%% assumptions -- compressibility effects, open edges. Here relaxation of
%% rigidity. Vs. elastic networks (momentum conserving, 3D mass
%% conserving).

%% Model: System definition. Assumptions: zero-velocity in xy (not
%% translation-invariant, held fixed in xy, but free in z
%% (motivation)). Deformation in z. Infinitely thick (much larger than
%% any r). Scales, dimensionless parameters: w, \tau_{in},
%% \tau_{el}. Simplicity: effective q2D theory valid for
%% r>>w. \sigma=1/2. Response to pointlike impulse (Green's function),
%% leading to two-point velocity correlations (FDT).

\section{Introduction}
%----------------------
\label{sec_intro}

Most fluids and soft materials around us\,---\,\eg suspensions, gels,
biomaterials\,---\,are complex, having intrinsic time and length
scales which are intermediate between the molecular and the
macroscopic ones \cite{WittenBook,DoiBook}. The standard method to
characterize the mechanical response of such materials, using
macroscopic rheometers \cite{LarsonBook}, has been supplemented in the
past two decades by {\it microrheology} \cite{MasonWeitz,review}. In
this technique the dynamics of tracer particles embedded in the
complex medium is used to infer the response, \ie the medium's
viscoelastic, frequency-dependent shear modulus
$G(\omega)$ \footnote{For brevity, as was done in earlier
  works, we use throughout the paper the same notation, $G(\omega)$,
  to denote both the complex frequency-dependent modulus, $G=G'+iG''$,
  and the one obtained from the real, time-dependent modulus $G(t)$
  through a one-sided Fourier (equivalently, Laplace) transform. The
  meaning in each instance should be clear from the context.}.  This
can be done either actively, driving the particles by an external
force, or passively, tracking their thermal fluctuations.

A particularly reliable variant of microrheology is {\it two-point
  microrheology} \cite{TwoPointCrocker,review}, in which one tracks
the correlated displacements of two particles as a function of their
mutual distance~$r$. Conservation of momentum in the medium guarantees
that, at sufficiently large $r$, velocity correlations between two
points in the medium must decay as $1/[G(\omega)r]$
\cite{Note1,HaimEPJE}. As a result, so do the correlations between the
displacements of two well-separated particles inside the
medium. Although it is insensitive to local details, unlike its
one-point counterpart \cite{TwoPointCrocker,Sonn2014,HaimEPJE},
two-point microrheology has not been widely used. The main reason is
that it requires a large amount of statistics to be acquired over a
limited time window, such that the mutual distance does not change
appreciably during the measurement. In addition, passive microrheology
is inapplicable in overly stiff media, where the thermal fluctuations
of the tracer particles are suppressed.

Suspensions confined between solid boundaries, as in microfluidic
channels, have been thoroughly studied in the past decade (see, \eg
Ref.~\cite{jpsj2009} and references therein). Of particular relevance
to the present work is the case of a quasi-two-dimensional (quasi-2D)
layer of colloid particles confined between two planar surfaces (see,
\eg \cite{prl2004HD,Santana2005,Bhattacharya2005,Alvarez2005}).  With
a few exceptions \cite{daddi2016hydrodynamic,daddi2016particle}, these
studies have considered confining surfaces that can be assumed
indefinitely rigid, such as glass plates. In this limit the fluid
loses its momentum through friction with the rigid boundaries and, as
a result, the correlations between confined particles are suppressed.
The suppression, however, is found to be weaker than what one would
expect\,---\,instead of the $1/r$ spatial decay mentioned above, the
pair-correlations in these quasi-2D suspensions decay as $1/r^2$. This
modified power law (rather than an exponential cutoff) originates in
the conservation of fluid mass \cite{prl2004HD,jpsj2009}.

One of our goals is to study what happens to the correlations in
confined fluids and suspensions once the infinite-rigidity assumption
is relaxed, and the boundaries are allowed to respond elastically or
viscoelastically. The second goal is to examine whether the
boundary-induced correlations could be used for a new type of
``noncontact'' two-point microrheology, where one would track the
correlated motions of particles immersed in the confined fluid rather
than the confining media. The concept of noncontact microrheology has
been introduced in the context of interfacial layers, using one-point
microrheology \cite{expDennin2013,expDennin2014} and
atomic force microscopy \cite{expAFM}. As will be shown below,
utilizing the distinct properties of quasi-2D suspensions confined
between two compliant plates should make noncontact microrheology
applicable to a broad class of soft materials.

%XXX
The article is organized as follows. After presenting the model in
Sec.~\ref{sec_model}, we divide the analysis into two parts. The first
(Sec.~\ref{sec_fluid}) concerns the flow response of the confined
fluid, in the absence of particles, to a localized impulse. In the
second part (Sec.~\ref{sec_particles}) we study the consequences of
this response for the displacement correlations between particles
embedded in the confined fluid. Finally, in Sec.~\ref{sec_discuss}, we
discuss the findings and their implications for noncontact two-point
microrheology.

\section{Model}
%--------------
\label{sec_model}

\begin{figure}[h]
  \centering
  \includegraphics[width=0.4\textwidth]{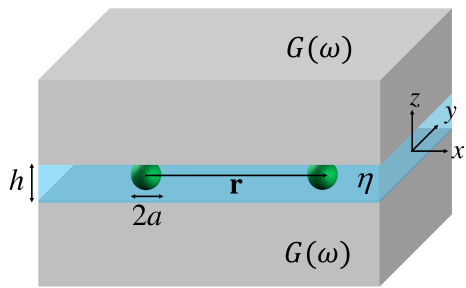}
  \caption{Schematic view of the system and its parameters.}
\label{fig_scheme}
\end{figure}

We consider a slit of thickness $h$, filled with an incompressible
viscous fluid of viscosity $\eta$. The slit is bounded by two
semi-infinite viscoelastic media, occupying the regions $z<-h/2$ and
$z>h/2$. A pair of particles of radius $a$ are embedded in the
confined fluid, their centers separated by the vector $\vecr$. See the
schematic illustration in Fig.~\ref{fig_scheme}. For simplicity we
assume that the viscoelastic media are identical, having the same
shear modulus $G(\omega)$, and that their frequency-dependent response
is fully captured by this modulus \cite{Note1} (in particular, inertia
is negligible). Two additional assumptions are employed. (a) We focus
on distances much larger than the slit thickness, $r\gg h>2a$. (b) The
flow of the confined fluid is assumed inertia-less (zero Reynolds
number). For the present system, the latter assumption implies that the
relaxation time required for fluid momentum to diffuse to the bounding
surfaces, $h^2\rho/\eta$ ($\rho$ being the fluid mass density), is
negligibly small compared to the examined time scale
$\omega^{-1}$. For water in a micron-wide slit this inertial
relaxation time is of the order of microseconds.

\section{Response of the confined fluid}
%---------------------------------------
\label{sec_fluid}

In the first part of the analysis we consider a particle-free confined
fluid and study its flow response to a point impulse.

%XXX
We begin with a heuristic account. Consider a localized oscillatory
force $F\cos(\omega t)$, applied at a point in the fluid. The
fluid-solid contact and the solid's response allow the stress to
propagate in three dimensions while conserving momentum. Imagine a
spherical envelope of radius $r$ around the force. For $r\gg h$ the
portion of envelope lying within the fluid-filled slit is
negligible. The solid's displacement far away then decays as $u\sim
F/(Gr)$ to guarantee that the stress (momentum flux) $\sigma\sim
G\nabla u\sim F/r^2$, once integrated over the envelope, equal the
momentum source $F$ for any choice of $r$. Back in the confined fluid,
due to the solid-fluid interface, this induces a flow velocity
$v\sim\omega u\sim \omega F/(Gr)$. Thus, at large distances, the
confined fluid has a velocity response that scales as $(\omega/G)/r$,
independent of the slit details. This response vanishes for an
infinitely rigid solid ($G\rightarrow\infty$)
%XXX
and under steady forcing
($\omega\rightarrow 0$).

We now turn to a more detailed analysis. The two assumptions mentioned
in Sec.~\ref{sec_model}\,---\,lateral distances much larger than the
slit thickness, and negligible relaxation time across that
thickness\,---\,allow us to use an effective 2D model for the confined
fluid. This so-called {\it lubrication approximation}
\cite{ReynoldsLubri, Oron1997} is obtained by integration of the
various variables over the thin dimension $z\in(-h/2,h/2)$. The
resulting reduced hydrodynamic equations are
\begin{eqnarray}
  && -\nabla p + \eta h \nabla^2 \vecv - \Gamma\left(\vecv -
  \dot{\vecu}\right) + \vecf = 0, 
\label{flow2D}\\ 
  && \nabla \cdot \vecv = 0,
\label{incompress2D}
\end{eqnarray}
expressing balance of forces (Eq.~(\ref{flow2D})) and
incompressibility (Eq.~(\ref{incompress2D})) of the quasi-2D flow. The
following 2D fields, which are functions of lateral position
$\vecr=(x,y)$ and time $t$, have been introduced: the flow velocity
$\vecv(\vecr,t)$, fluid pressure $p(\vecr,t)$, surface displacement of
the confining media at the slit boundaries $\vecu(\vecr,t)$, and an
external force density $\vecf(\vecr,t)$, applied on the fluid parallel
to the $(x,y)$ plane. The symbols $\nabla$ and $(\,\dot{}\,)$ denote a
2D gradient and a time derivative.

%XXX
A friction term, with a friction coefficient $\Gamma$, appears in the
force balance, Eq.~(\ref{flow2D}), characterizing the momentum
exchange between the fluid and the confining media. To keep the
analysis as general as possible, we do not specify $\Gamma$. This
parameter depends on details of the flow profile in the $z$ direction,
created in the slit by $\vecf$, in particular, the boundary conditions
at the fluid-solid interfaces. For example, in the simplest case of
static no-slip boundary conditions at flat surfaces and a parabolic
velocity profile (Poiseuille flow), we have $\Gamma=\eta/(12h)$
\cite{HB}.  If out-of-plane boundary deformation is included in these
boundary conditions, the integration over the thin dimension may yield
$\Gamma$ which is also time-dependent, turning the friction term in
Eq.~(\ref{flow2D}) into a convolution over time. From now on we
Fourier-transform all functions from the time domain to the frequency
domain, $t\rightarrow\omega$. Accordingly, the coefficient $\Gamma$
appearing in the derivation below can be considered a
frequency-dependent function, $\Gamma(\omega)$, without loss of
generality.

Our aim now is to find the velocity response of the quasi-2D fluid,
i.e., the Green's function $\tencG^f(\vecr,\omega)$ connecting the
force density at point $\vecr'$ with the flow velocity at point
$\vecr$ according to $v_i(\vecr,\omega) = \int d^2r'
\cG^f_{ij}(\vecr-\vecr',\omega)f_j(\vecr',\omega)$. To calculate it we
need the solid's response as well, since the two are coupled. [See the
  frictional coupling to $\vecu$ in Eq.~(\ref{flow2D}).]
%XXX
The calculation is done through the following sequence of five logical
steps. (i) A 2D force density $\vecf$ is applied in the fluid. (ii) It
results in the yet-unknown 2D flow velocity $\vecv$ (the goal of the
calculation) following Eqs.~(\ref{flow2D}) and (\ref{incompress2D}).
(iii) Given a yet-unknown surface displacement $\vecu$, a friction
force per unit area, $\Gamma(\vecv-\dot{\vecu})$, is applied on
the surfaces. (iv) The solid responds to this surface force by a 3D
displacement field, whose value at the surface is $\vecu$; thus, once
the solid response has been dealt with, one obtains a relation between
$\vecv$ and $\vecu$. (v) This relation is substituted back in the flow
equations (\ref{flow2D}) and (\ref{incompress2D}) to obtain $\vecv$ (and
$\vecu$ if so desired).

To account for the solid's response, in principle, one should solve
the appropriate viscoelastic equations, along with boundary conditions
at the slit surfaces which will ensure the continuity of stress across
the boundaries. However, the corresponding 3D equations and boundary
conditions can both be bypassed using a Green's function formulation,
%XXX
which will connect the flow-induced friction force density mentioned
above, $\vecP(\vecr',\omega)$, acting on each of the solid surfaces at
some point $\vecr'$, with the surface displacement at point $\vecr$,
\begin{equation}
  u_i(\vecr,\omega) = \int d^2r' \cG^s_{ij}(\vecr-\vecr',\omega)
  P_j(\vecr',\omega).
\label{GreenSolid}
\end{equation}
This formulation allows us also to remain within the 2D-reduced
description. The 3D Green's function for a point force exerted on the
surface of a solid is calculated in Ref.~\cite{LLelasticity}. Once
specialized to forces and surface displacements parallel to the
surface, it reduces to
\begin{equation}
  \cG^s_{ij}(\vecr,\omega) = \frac{1-\nu}{2\pi
    G(\omega)r}\left(\delta_{ij} + \frac{\nu}{1-\nu}\,
  \frac{r_ir_j}{r^2} \right),
\label{Gsr}
\end{equation}
where $\nu$ is the solid's Poisson ratio. In 2D Fourier space,
$\vecr\rightarrow\vecq$,
\begin{equation}
  \cG^s_{ij}(\vecq,\omega) = \frac{1}{G(\omega)q}\left(\delta_{ij} - \nu\,
  \frac{q_iq_j}{q^2} \right).
\label{Gsq}
\end{equation}
For Eqs.~(\ref{Gsr}) and (\ref{Gsq}) to properly describe a solid, we
assume a finite shear modulus at steady state, $G(\omega=0)>0$.

In the 2D-reduced description, the only force exerted on the solid
surfaces is the flow-induced friction appearing in
Eq.~(\ref{flow2D}). Thus, the force per unit area exerted on each of
the two surfaces is
\begin{equation}
  \vecP=-\frac{1}{2} \Gamma(i\omega\vecu - \vecv).
\end{equation}
Using this surface force in Eqs.~(\ref{flow2D})--(\ref{Gsq}) and solving for
$\vecv$ while considering a localized impulse in the fluid,
$\vecf(\vecr,t)=\delta(\vecr)\delta(t)$, we obtain the following
expression for the fluid velocity response:
\begin{equation}
    \cG^f_{ij}(\vecq,\omega) = \frac{Gq+i\Gamma\omega/2}{q[ G\eta hq^2 + 
      \Gamma(G + i\eta\omega hq/2)]} \left( \delta_{ij}
  - \frac{q_iq_j}{q^2} \right).
\label{Gfq}
\end{equation}
In the limit of $G\rightarrow\infty$ (infinitely rigid solid), as well
as in the limit of $\omega\rightarrow 0$ (steady state),
Eq.~(\ref{Gfq}) reduces to the known large-distance response function
of a fluid between two infinitely rigid walls \cite{jpsj2009},
$\cG^f_{ij}(\vecq) = (\eta hq^2+\Gamma)^{-1}(\delta_{ij} -
\hat{q}_i\hat{q}_j)$. Another instructive limit is
$\Gamma\rightarrow\infty$, forcing the bounding surfaces to move
together with the fluid. In this limit Eq.~(\ref{Gfq}) becomes
$\cG^f_{ij}(\vecq) = (\eta h q^2+2q G/i\omega)^{-1} (\delta_{ij} -
\hat{q}_i\hat{q}_j)$, reproducing the known result for a membrane (of
2D viscosity $\eta h$) embedded in a liquid (of viscosity $G/i\omega$)
\cite{Komura2012}. Equation (\ref{Gfq}) can be inverted back to real
space, yielding a complicated expression. In the limit of $r\gg h$
(which is anyhow demanded by the theory), we get
\begin{eqnarray}
  \cG^f_{ij}(\vecr,\omega) =  &&-\frac{1}{2\pi\Gamma r^2} \left(1 + 
  \frac{\Gamma \eta h\omega^2}{4G^2} \right) \left( \delta_{ij}
  - 2\frac{r_ir_j}{r^2} \right) \nonumber\\
  &&+ \frac{i\omega}{4\pi G r}\,\frac{r_ir_j}{r^2}.
\label{Gfr}
\end{eqnarray}

\begin{figure}[ht]
  \centerline{{\large ({\it a})}~\includegraphics[width=0.3\textwidth]{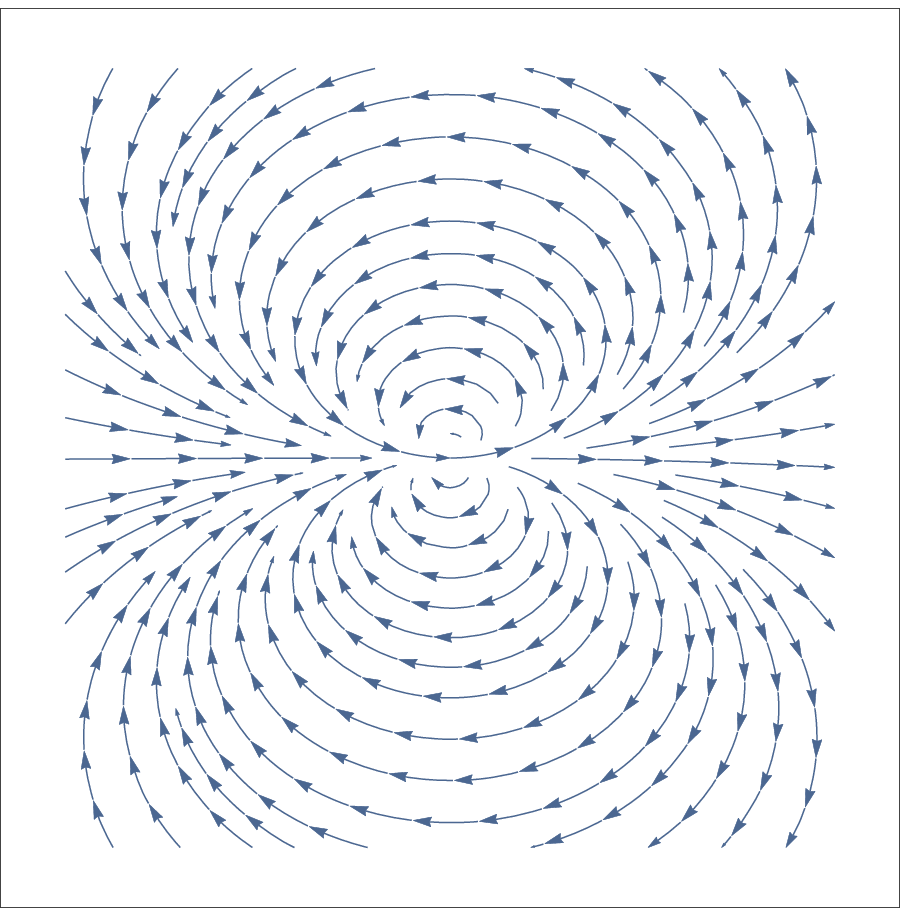}}
  \centerline{{\large ({\it b})}~\includegraphics[width=0.3\textwidth]{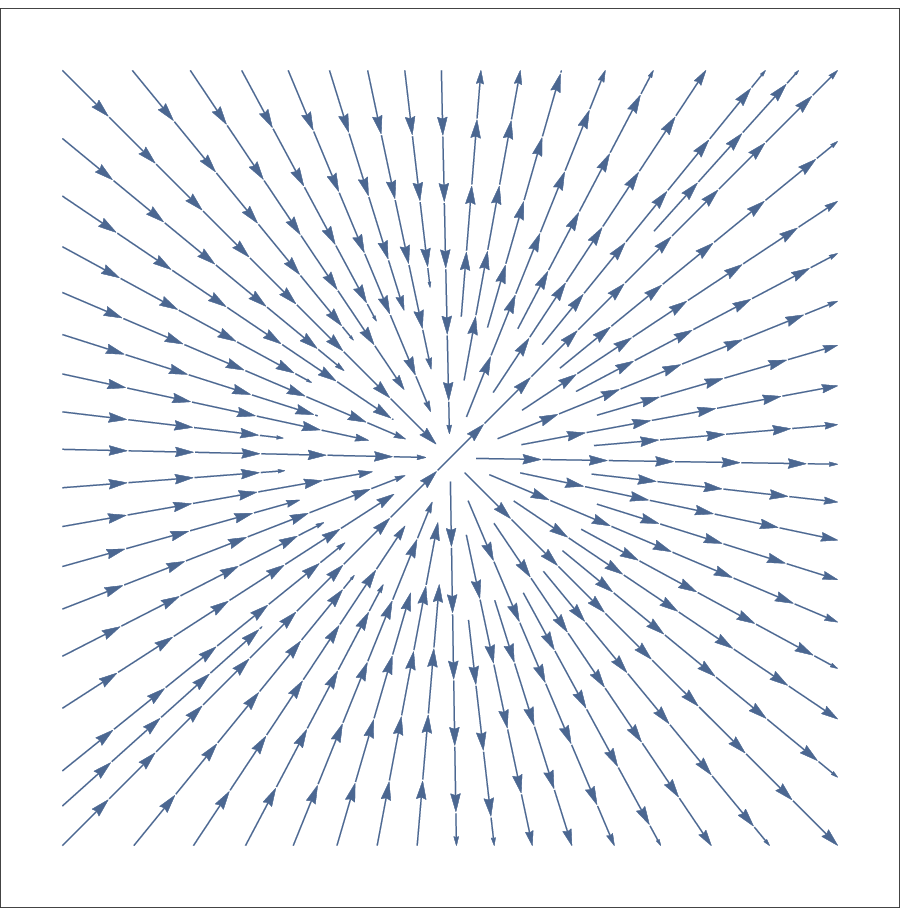}}
  \centerline{{\large ({\it c})}~\includegraphics[width=0.3\textwidth]{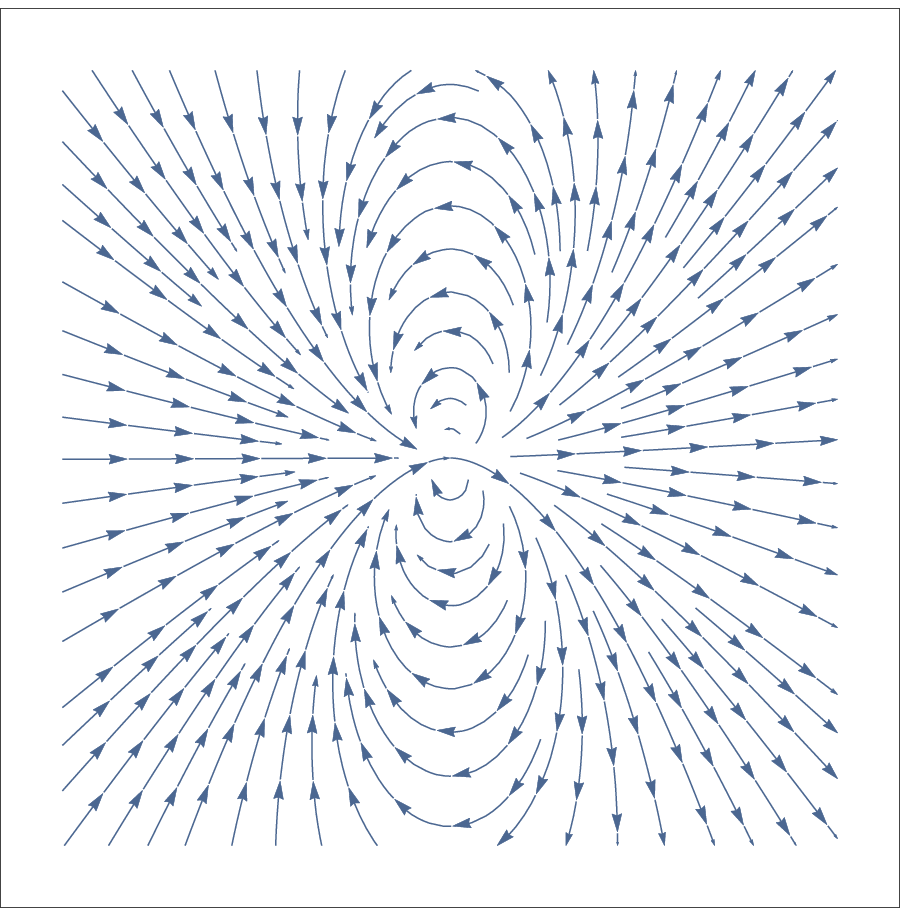}}
  \caption{2D flow fields in the confined fluid, arising from a point impulse,  $\vecf=f\xhat$. (a) Monopolar term. (b) Dipolar term. (c) Total flow field.}
  \label{fig_flowfields}
\end{figure}

Equation (\ref{Gfr}), describing the spatio-temporal response of the
quasi-2D fluid, is one of our central results. Its two terms reflect
the two conservation laws at play. The first, decaying as $1/r^2$ and
having a 2D-dipolar form, arises from mass conservation of the fluid,
i.e., the response to the effective 2D mass dipole created by the
confined localized force \cite{jpsj2009}. The flow lines produced by this
term are depicted in Fig.\ \ref{fig_flowfields}(a). The second term,
decaying as $1/r$, reflects momentum conservation of the entire
system, i.e., the response to the 3D momentum monopole created by
$\vecf$. The angular form of this term is dictated by the
incompressibility of the flow, $\pd_i(r_ir_j/r^3)=0$.  The
corresponding flow field is shown in Fig.~
\ref{fig_flowfields}(b). Figure~\ref{fig_flowfields}(c) shows the
combination of the two flow fields, with the dipolar component
dominating at shorter distances (yet larger than $h$), and the
monopolar one being dominant at the larger distances. The crossover between
these two regions occurs around the frequency-dependent distance
$\lc\sim G/(\omega\Gamma) \sim [G/(\eta\omega)]h$. With decreasing
frequency, the crossover distance increases until, at steady state
($\omega=0$), it becomes indefinitely large. In this long-time limit
we are left with the mass-dipole response alone, as is known for
quasi-2D fluids confined between infinitely rigid surfaces
\cite{prl2004HD}. More precisely, this limit is obtained for
$\omega\ll G/\eta$; for such low frequencies the solid deformation
rate is too small to significantly affect the flow. Conversely, for
frequencies higher than this value, the viscoelastic contributions to
the fluid response become significant. Apart from the crossover
arising from the last term of Eq.~(\ref{Gfr}), the equation contains a
smaller viscoelastic correction to the $1/r^2$ term, of order
$(\omega\eta/G)^2$, which is of lesser significance for the following
discussion.

Equation (\ref{Gfr}) also gives immediately, through
the fluctuation-dissipation theorem, the cross-correlations between
the flow velocities at two points separated by $\vecr$,
\begin{equation}
  C_{ij}(\vecr,\omega) \equiv \langle v_i({\bf
    0},\omega)v_j(\vecr,-\omega)\rangle = \kT
  \cG^f_{ij}(\vecr,\omega),
\label{vv}
\end{equation}
where $\kT$ is the thermal energy.

\section{Displacement correlations of embedded particles}
%--------------------------------------------------------
\label{sec_particles}

We now turn to the consequences of the fluid response found in the
preceding section for the dynamics of particles embedded in the
fluid. In two-point microrheology one measures the cross-correlations
between the displacements of two tracer particles,
$\Delta\vecR^{(1)}(t)$ and $\Delta\vecR^{(2)}(t)$, whose positions are
separated by $\vecr$, during the time interval $t$,
$D_{ij}(\vecr,t)\equiv \langle\Delta R^{(1)}_i(t)\Delta
R^{(2)}_j(t)\rangle$. They are conventionally decomposed into a
longitudinal correlation and a transverse one, where the displacements
are projected on, and perpendicular to, the separation vector
$\vecr$. Assuming that the pair separation is much larger than the
particle radius $a$, $r\gg a$, the particles' displacement
correlations are obtained from the fluid's velocity correlations
through a double time integration, or, in the frequency domain,
\begin{eqnarray}
  D_\parallel(r,\omega) &=& -\frac{2}{\omega^2} C_{xx}(r\xhat,\omega)
  \label{Dparallel}\\
  &=& -\frac{k_{\rm B}T\alpha h}{\pi\eta\omega^2} \left(1 + 
  \frac{\eta^2\omega^2}{4\alpha G^2}\right) \frac{1}{r^2}
  + \frac{k_{\rm B}T}{2\pi i\omega G} \frac{1}{r}, \nonumber\\
  D_\perp(r,\omega) &=& -\frac{2}{\omega^2} C_{xx}(r\yhat,\omega)
  \label{Dperp}\\
  &=&  \frac{k_{\rm B}T\alpha h}{\pi\eta\omega^2} \left(1 + 
  \frac{\eta^2\omega^2}{4\alpha G^2} \right) \frac{1}{r^2}. \nonumber
\end{eqnarray}
In Eqs.~(\ref{Dparallel}) and (\ref{Dperp}) we have introduced a
dimensionless coefficient, $\alpha$, relating the friction coefficient
$\Gamma$ to the actual system's parameters according to
$\Gamma^{-1}=\alpha h/\eta$. It depends on the degree of particle
confinement, $a/h$, and the boundary conditions at the fluid-solid
interfaces. Once again, we prefer to remain on the most general level
and regard it as a phenomenological parameter to be determined by
experiment or a more detailed calculation for a specific system. We
note, however, that the value of $\alpha$ is typically small and
depends weakly on $a/h$. For pointlike particles and no-slip boundary
conditions, one has $\alpha(a/h\rightarrow 0)\simeq 0.030$
\cite{LironMochon}, decreasing for larger confinement ratio to
$\alpha(a/h\simeq 0.45)\simeq 0.019$ \cite{HD2005}. (For no-slip
boundary conditions $\alpha$ vanishes for $a/h=1/2$, as the sphere
touches the walls.)

\begin{figure}[h]
	\centering \includegraphics[width=0.43\textwidth]{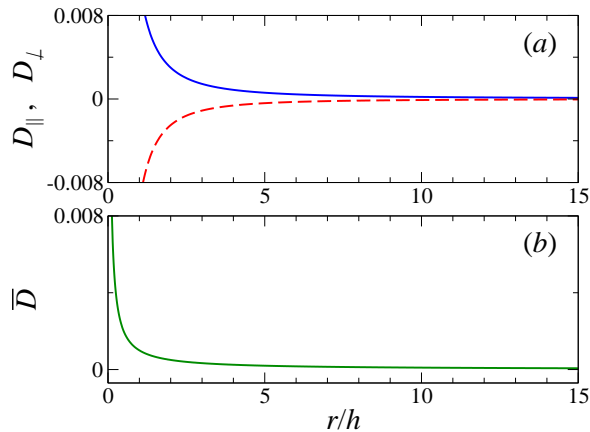}
        \caption{Two-point displacement correlations, normalized by
          $-\pi\omega^2\eta h/\kT$, as a function of particle
          separation, normalized by $h$. Once normalized, these
          functions depend only on the parameters $G/(\eta\omega)$ and
          $\alpha$.  We have used $G/(\eta\omega)=500$ and
          $\alpha=0.01$. Panel (a) shows the longitudinal (solid
          blue) and transverse (dashed red) correlations. Panel (b)
          presents the residual correlation obtained by adding together
          the two correlations of (a).}
	\label{fig_disp_correl}
\end{figure}

Equations (\ref{Dparallel}) and (\ref{Dperp}) contain our main
experimental predictions. They can be used to extract the viscoelastic
shear modulus $G(\omega)$ of the confining media from the displacement
correlations of particles lying {\em outside of them}, inside the
confined viscous fluid. Since only shear stresses are exerted on the
boundaries, Eqs.~(\ref{Dparallel}) and (\ref{Dperp}) are independent
of the Poisson ratio and thus cannot be used to measure it. (The main
ingredient of most soft materials, anyway, is a molecular solvent,
making them nearly incompressible, with $\nu\simeq 1/2$.) The
dependencies of the longitudinal and transverse displacement
correlations on the distance between the two particles (for a given
frequency) are presented in Fig.~\ref{fig_disp_correl}(a). As seen in
Eqs.~(\ref{Dparallel}) and (\ref{Dperp}), and as was observed
experimentally for rigid plates \cite{prl2004HD}, the $1/r^2$ terms
have opposite signs in the two correlations. This is a result of the
dipolar shape of the underlying response (see
Fig.~\ref{fig_flowfields}(a)). While the transverse correlation
contains only this contribution for $r\gg h$, the longitudinal one
crosses over around $\lc\sim [\alpha G/(\eta\omega)]h$ to the $1/r$
monopolar term. (As noted above, the absence of the $1/r$ term in
$D_\perp$ is dictated by the incompressibility of the flow.)
Moreover, the particular spatial decays of the two correlations can be
exploited to isolate the monopolar, viscoelastic contribution,
\begin{equation}
  \Dbar(r,\omega) \equiv D_\parallel(r,\omega) + D_\perp(r,\omega) =
  \frac{k_{\rm B}T}{2\pi i\omega G} \frac{1}{r}.
\label{monopolar}
\end{equation}
This residual pair-correlation is shown in
Fig.~\ref{fig_disp_correl}(b). Equation~(\ref{monopolar}) might be the
most useful result of this work. Adding up the two measured
correlations is predicted to leave a residual long-range correlation,
$\Dbar \sim 1/r$, whose coefficient is inversely proportional to
$G(\omega)$.

\section{Discussion}
%-------------------
\label{sec_discuss}

From the fundamental point of view, our findings indicate the
existence of long-ranged correlations in confined fluids and between
particles suspended in them. Not only do the steady-state correlations
decay as $1/r^2$, as was found earlier for rigid slits
\cite{prl2004HD,Santana2005,Bhattacharya2005,Alvarez2005}, the
compliance of the confining media leads to an even weaker spatial
decay of the time-dependent correlations, falling off as $1/r$. In
addition to the pair-correlations analyzed here, these long-ranged
effects should influence the collective dynamics of suspensions
confined in soft materials, which calls for further study.

From the practical perspective, the measurable correlation defined in
Eq.~(\ref{monopolar}) suggests a noncontact, two-point,
microrheological technique of several key advantages. (i) While the
signal $\Dbar$ should not be weaker than the one used in standard
two-point microrheology, which also scales as $\kT/(\omega G r)$
\cite{TwoPointCrocker}, it is extracted from the much stronger
fluctuations of particles suspended in a viscous fluid. (ii) Arising
from the conservation of momentum in the entire 3D system, it is
guaranteed to be the only correlation term remaining at sufficiently
large distances.  (iii) For the same reason, it is independent of
details of the confining slit, such as its width $h$ and coefficient
$\alpha$. Thus, in principle, one does not have to measure these
parameters before using $\Dbar$ to extract $G$. (iv) Furthermore, the
residual correlation should be robust also against modifications of
the confining surfaces; one can coat the surfaces with thin layers of
different composition so long as their thickness is much smaller than
the sampled lateral distances. (v) All particles in the slit can be
imaged on the focal plane. (vi) Since large-distance pair correlations
in the confined suspension are practically independent of
concentration \cite{HD2005}, one can use dense suspensions to improve
the statistics. Indeed, two-point correlations in quasi-2D suspensions
confined between two rigid surfaces have been measured with remarkable
accuracy over distances more than ten times larger than the
confinement width $h$ \cite{prl2004HD,HD2005}.

At the same time, in practice, one cannot expect exact mutual
cancellation of the $\pm 1/r^2$ terms in the measured $D_\parallel$
and $D_\perp$. For example, assuming noise of a few percent in the
measurements, we need the $1/r$ term to be larger than a few percent
of the $1/r^2$ term. This will occur at distances $r/h\gtrsim
10^{-2}\alpha[G/(\omega\eta)]$. An entangled F-actin network has
$G/\omega\sim 1$~Pa$\cdot$s over a frequency range of $1$--$100$~Hz
\cite{Sonn2014}, yielding for confined water
($\eta=10^{-3}$~Pa$\cdot$s) a strong signal at all relevant
distances. (Note that the small value of $\alpha$ helps the dominance
of the $1/r$ signal.) In the case of stiffer materials, such as
typical elastomers ($G\sim 1$~MPa), one would need higher frequencies,
$\omega \gtrsim 1$~kHz (or time scales of order milliseconds), to
observe a similar effect. Another experimental restriction is that the
confining media should be much thicker than the largest lateral
distance between tracked particles. Overall, for soft biomaterials,
the main complication in applying the noncontact microrheology
proposed here is expected to be the preparation of the required
double-slab sample.

\begin{acknowledgments}
We are grateful to Shigeyuki Komura for spotting an error in an
earlier version of the article. We thank Yael Roichman for helpful
discussions. This work has been supported by the Israel Science
Foundation (Grant No.\ 164/14).
\end{acknowledgments}

\bibliography{pre172_bibl}

\end{document}